\documentclass[a4paper,11pt]{article}

\usepackage{jheppub} 

\usepackage[T1]{fontenc} 
\usepackage{lineno}

\newcommand{\affuni}[2]{Dipartimento di Fisica dell'Universit\`a #1, #2, Italy.}
\newcommand{\affinfn}[2]{INFN Sezione di #1, #2, Italy.}

\title{\boldmath Measurement of $\eta$ meson production in $\gamma \gamma$ interactions 
and $\Gamma(\eta \to \gamma \gamma)$ with the KLOE detector}

\collaboration{The KLOE-2 Collaboration}
\author[h]{D.~Babusci,}
\author[r,s]{D.~Badoni,}
\author[g]{I.~Balwierz-Pytko,}
\author[h]{G.~Bencivenni,}
\author[p,q]{C.~Bini,}
\author[h]{C.~Bloise,}
\author[h]{F.~Bossi,}
\author[u]{P.~Branchini,}
\author[t,u]{A.~Budano,}
\author[w]{L.~Caldeira~Balkest\aa hl,}
\author[h]{G.~Capon,}
\author[t,u]{F.~Ceradini,} 
\author[h]{P.~Ciambrone,}
\author[g]{E.~Czerwi\'nski,}
\author[h]{E.~Dan\`e,}
\author[h]{E.~De~Lucia,}
\author[b]{G.~De~Robertis,}
\author[p,q]{A.~De~Santis,}
\author[p,q]{A.~Di~Domenico,}
\author[l,m]{C.~Di~Donato,}
\author[s]{R.~Di~Salvo,}
\author[h]{D.~Domenici,}
\author[a,b]{O.~Erriquez,}
\author[a,b]{G.~Fanizzi,}
\author[r,s]{A.~Fantini,}
\author[h]{G.~Felici,}
\author[p,q]{S.~Fiore,}
\author[p,q]{P.~Franzini,}
\author[p,q]{P.~Gauzzi,}
\author[j,d]{G.~Giardina,}
\author[h]{S.~Giovannella,}
\author[r,s]{F.~Gonnella,}
\author[u]{E.~Graziani,}
\author[h]{F.~Happacher,}
\author[w]{L.~Heijkenskj\"old,}
\author[w]{B.~H\"oistad,}
\author[h]{L.~Iafolla,}
\author[w]{M.~Jacewicz,}
\author[w]{T.~Johansson,}
\author[w]{A.~Kupsc,}
\author[h,v]{J.~Lee-Franzini,}
\author[h]{B.~Leverington,}
\author[b]{F.~Loddo,}
\author[t,u]{S.~Loffredo,}
\author[j,d,c]{G.~Mandaglio,}
\author[k]{M.~Martemianov,}
\author[h,o]{M.~Martini,}
\author[r,s]{M.~Mascolo,}
\author[r,s]{R.~Messi,}
\author[h]{S.~Miscetti,}
\author[h]{G.~Morello,}
\author[s]{D.~Moricciani,}
\author[g]{P.~Moskal,}
\author[u,y,1]{F.~Nguyen,%
\note{corresponding authors}}
\author[u]{A.~Passeri,}
\author[n,h]{V.~Patera,}
\author[t,u]{I.~Prado~Longhi,}
\author[b]{A.~Ranieri,}
\author[i]{C.~F.~Redmer,}
\author[h]{P.~Santangelo,}
\author[h]{I.~Sarra,}
\author[e,f]{M.~Schioppa,}
\author[h]{B.~Sciascia,}
\author[g]{M.~Silarski,}
\author[t,u,1]{C.~Taccini,}
\author[u]{L.~Tortora,}
\author[h]{G.~Venanzoni,}
\author[x]{W.~Wi\'slicki,}
\author[w]{M.~Wolke}
\author[g]{and J.~Zdebik}
\affiliation[a]{\affuni{di Bari}{Bari}}
\affiliation[b]{\affinfn{Bari}{Bari}}
\affiliation[c]{Centro Siciliano di Fisica Nucleare e Struttura della Materia, Catania, Italy.}
\affiliation[d]{\affinfn{Catania}{Catania}}
\affiliation[e]{\affuni{della Calabria}{Cosenza}}
\affiliation[f]{INFN Gruppo collegato di Cosenza, Cosenza, Italy.}
\affiliation[g]{Institute of Physics, Jagiellonian University, Cracow, Poland.}
\affiliation[h]{Laboratori Nazionali di Frascati dell'INFN, Frascati, Italy.}
\affiliation[i]{Institut f\"ur Kernphysik, 
Johannes Gutenberg Universit\"at Mainz, Germany.}
\affiliation[j]{Dipartimento di Fisica e Scienze della Terra dell'Universit\`a di Messina, Messina, Italy.}
\affiliation[k]{Institute for Theoretical and Experimental Physics (ITEP), Moscow, Russia.}
\affiliation[l]{\affuni{``Federico II''}{Napoli}}
\affiliation[m]{\affinfn{Napoli}{Napoli}}
\affiliation[n]{Dipartimento di Scienze di Base ed Applicate per l'Ingegneria dell'Universit\`a 
``Sapienza'', Roma, Italy.}
\affiliation[o]{Dipartimento di Scienze e Tecnologie applicate, Universit\`a ``Guglielmo Marconi", Roma, Italy.}
\affiliation[p]{\affuni{``Sapienza''}{Roma}}
\affiliation[q]{\affinfn{Roma}{Roma}}
\affiliation[r]{\affuni{``Tor Vergata''}{Roma}}
\affiliation[s]{\affinfn{Roma Tor Vergata}{Roma}}
\affiliation[t]{\affuni{``Roma Tre''}{Roma}}
\affiliation[u]{\affinfn{Roma Tre}{Roma}}
\affiliation[v]{Physics Department, State University of New 
York at Stony Brook, USA.}
\affiliation[w]{Department of Physics and Astronomy, Uppsala University, Uppsala, Sweden.}
\affiliation[x]{National Centre for Nuclear Research, Warsaw, Poland.}
\affiliation[y]{Present Address: Laborat\'orio de Instrumenta\c{c}\~{a}o e F\'isica Experimental de Part\'iculas,
Lisbon, Portugal.}

\emailAdd{federico.nguyen@cern.ch}
\emailAdd{cecilia.taccini@roma3.infn.it}


\abstract{We present a measurement of $\eta$ meson production in photon-photon interactions produced by electron-positron beams colliding with $\sqrt{s} = 1$ GeV. The measurement is done with the KLOE detector at the $\phi$-factory DA$\Phi$NE with an integrated luminosity of 0.24 fb$^{-1}$. The $e^+ e^- \to e^+ e^- \eta$ cross section is measured without detecting the outgoing electron and positron, selecting the decays $\eta \to \pi^+ \pi^- \pi^0$ and $\eta \to \pi^0 \pi^0 \pi^0$. The most relevant background is due to $e^+ e^- \to \eta \gamma$ when the monochromatic photon escapes detection. 
The cross section for this process is measured as $\sigma(e^+ e^- \to \eta \gamma) = (856 \pm 8_{\mbox{stat}} \pm 16_{\mbox{syst}})$ pb. 
The combined result for the $e^+ e^- \to e^+ e^- \eta$ cross section is $\sigma(e^+ e^- \to e^+ e^- \eta) = (32.72 \pm 1.27_{\mbox{stat}} \pm 0.70_{\mbox{syst}})$ pb. From this we derive the partial width $\Gamma(\eta \to \gamma \gamma) = ( 520 \pm 20_{\mbox{stat}} \pm 13_{\mbox{syst}} )$ eV. This is in agreement with the world average and is the most precise measurement to date.}

\begin{document}
\maketitle
\flushbottom

\section{Introduction} 
\label{s:INTRODUCTION}
Photon-photon production of neutral mesons provides basic information on their structure. 
The strength of the coupling, measured by the partial decay width $\Gamma(X \to \gamma\gamma)$, is related to the quark content of the meson and gives information on the relations between the hadronic state and its $q\bar{q}$ representation. For the light pseudoscalar mesons $\pi^0$, $\eta$ and $\eta'$, the coupling to real photons is measured in their $\gamma\gamma$ decays, while the coupling to space-like photons can be measured in $\gamma\gamma$ interactions. This is of particular interest in evaluating the light-by-light contribution to the anomalous magnetic moment of the muon~\cite{MUONg-2}.
Photon-photon interactions in electron-positron colliders were pioneered at the Frascati collider Adone in the $'70$s~\cite{ggADONE1,ggADONE2,ggADONE3} and since then have been used to study the production of hadrons in almost all $e^+ e^-$ colliders in a variety of conditions in low- and high-$q^2$ processes~\cite{ggREVIEW1,ggREVIEW2,ggREVIEW3}. In particular, measurements of the $\gamma \gamma$ partial width of $\eta$ and $\eta'$ mesons have been done measuring the  $e^+ e^- \to e^+ e^- \eta (\eta')$ cross section~\cite{GAMMAETA1a,GAMMAETA1b,GAMMAETA2,GAMMAETA3,GAMMAETA4,GAMMAETA5}. 

We present a measurement of the cross section $e^+ e^- \to e^+ e^- \eta$ with the KLOE detector at the $\phi$-factory DA$\Phi$NE. The cross section $\sigma(e^+ e^- \to e^+ e^- \eta)$ is  a convolution of the differential $\gamma \gamma$ luminosity and the $\gamma \gamma \to \eta$ cross section. The $\eta$ partial decay width $\Gamma(\eta \to \gamma \gamma)$ is obtained by extrapolating the value of $\sigma(\gamma \gamma \to \eta)$ for real photons.

DA$\Phi$NE is an $e^+ e^-$ collider designed to operate at high luminosity at the mass of the $\phi$ resonance, 1020 MeV. We  analyzed data collected with DA$\Phi$NE operating off the $\phi$ peak, at $\sqrt{s} = 1$ GeV, to reduce the large background  from $\phi$ decays. The final state $e^+$ and $e^-$ are not detected, being emitted with high probability in the forward directions outside the acceptance of the detector. The production of the $\eta$ meson is identified in two decay modes, $\eta \to \pi^+ \pi^- \pi^0$ and $\eta \to \pi^0 \pi^0 \pi^0$, that exploit in a complementary way the tracking system and calorimeter of the detector. The most relevant background is the radiative process $e^+ e^- \to \eta \gamma$ and, in both measurements, the yield of $\eta$ mesons is controlled by the $e^+ e^- \to \eta \gamma$ cross section measured in the same data sample with a dedicated analysis. The data sample used in the analyses corresponds to an integrated luminosity of 0.24 fb$^{-1}$. \\

\section{Signal and background model} 
\label{s:MODEL}
For electron and positron beams colliding with energy $E$, the cross section
for production of a state $X$ in $\gamma \gamma$ interactions
with photon 4-momenta $q_{1}$ and $q_{2}$ is
\begin{equation}
\sigma(e^{+} e^{-} \to e^{+} e^{-} X) = \int \sigma_{\gamma \gamma \to X}(q_1, q_2)\ 
\Phi(q_1, q_2)\ \frac{d\vec{q}_1}{E_1} \frac{d\vec{q}_2}{E_2} \,,
\label{eq:eeXSECTION}
\end{equation}
where the $\gamma \gamma$ differential luminosity $\Phi(q_1, q_2)$ has been calculated 
in~\cite{BKT,BBM,Budnev} using different approximations and is proportional to $(\alpha/2 \pi)^2 (\ln E/m_e)^2$.
For a narrow resonance of spin 0 the formation cross section is
\begin{equation}
 \sigma_{\gamma \gamma \to X} = \frac{8 \pi^2}{m_X} \Gamma_{X \to \gamma \gamma}\ 
 \delta(w^2 - m^2_X)\ |F(q_1^2,q_2^2)|^2\,,
  \label{eq:ggXSECTION}
\end{equation}
where $\Gamma_{X \to \gamma \gamma}$ is the radiative width, and $w^2 = (q_1 + q_2)^2$. The transition form factor, $F(q_1^2,q_2^2)$, is equal to one for real photons and is usually parametrized in the form
\begin{equation}
 F(q_1^2,q_2^2) = \frac{1}{1 - b q_1^2}\ \frac{1}{1 - b q_2^2} \,,
  \label{eq:TRANSFF}
\end{equation}
inspired by the Vector Dominance Model~\cite{VDM}. 
The parameter $b$ for the $\eta$ meson has been measured at high $q^2$ values in $\gamma \gamma$ 
experiments with single-tagging~\cite{bTPC,bCELLO,bCLEO} and in the $\eta$ leptonic radiative 
decays $\eta \to \ell^+ \ell^- \gamma$~\cite{LeptonG,NA60,MAMI} at low $q^2$ values, closer to those of this measurement. 
The results do not show appreciable dependence on $q^2$ and the value assumed in this analysis, 
$b_{\eta} = (1.94 \pm 0.15)$ GeV$^{-2}$, was obtained as an average of the measurements at low $q^2$. 

The detector response for signal and background events is fully
simulated with the Monte Carlo (MC) program \texttt{Geanfi}
\cite{K:FILFO}. While \texttt{Geanfi} contains the event
generator for all background processes, a new generator for $e^{+}
e^{-} \to e^{+} e^{-} X$ events is developed and interfaced to
the detector simulation.
Events are generated with exact matrix element according to full 3-body phase space
distributions~\cite{NPP:sigma}. This results
in the production of $\eta$ mesons with non negligible transverse momentum. The relative
error due to high-order radiative corrections to equation (\ref{eq:eeXSECTION}) is estimated to be 1\%~\cite{teo}. 
All background processes have been extensively 
studied in other analyses. A source of irreducible background is the reaction $e^+ e^- \to \eta \gamma$
when the monochromatic photon is emitted at small angles and is not detected. The cross section for 
this process is measured in the same data sample 
with two independent methods and the results agree with each other providing an important consistency check of the analysis. The beam-induced backgrounds were measured during data taking
and background events are added to simulated events in the MC on a run-by-run basis.  

\section{The KLOE detector} 
\label{s:DETECTOR}
The KLOE detector consists of a large volume cylindrical drift chamber, surrounded by
a lead-scintillating fibers finely segmented calorimeter. 
A superconducting coil around the calorimeter 
provides a 0.52 T axial magnetic field. The beam pipe at the interaction region is 
spherical in shape with 10 cm radius, it is made of a Beryllium--Aluminum alloy of 
0.5 mm thickness. Low-beta quadrupoles are located at $\sim50$ cm distance from 
the interaction region. Two small lead-scintillating tiles calorimeters (QCAL)~\cite{K:qcal} 
are wrapped around the quadrupoles.

The drift chamber (DC)~\cite{K:drift}, 4 m in diameter and 3.3 m long, has 12,582 drift cells 
arranged in 58 concentric rings with alternated stereo angles and is filled with a 
low-density gas mixture of 90\% Helium--10\% isobutane. The chamber shell is made of 
carbon fiber-epoxy composite with an internal wall of 1.1 mm thickness at 25 cm radius. 
The spatial resolutions are $\sigma_{xy} \sim150\ \mu$m and $\sigma_{z} \sim2$ mm 
\footnote{\ KLOE uses a coordinate system where $z$ is the bisector 
of the electron and positron beams, $x$ and $y$ define the transverse plane.}. 
The momentum resolution for long tracks is $\sigma(p_T)/p_T \sim0.4\%$. 
Vertices are reconstructed with a spatial resolution of $\sim3$ mm. 

The calorimeter~\cite{K:calo} is divided into a barrel and two end-caps and covers 98\% 
of the solid angle. The readout granularity is $(4.4\times4.4)$ cm$^2$, 
for a total of 2440 cells arranged in five layers. Each cell is read out at both ends by photomultipliers. 
The energy deposits are obtained 
from the signal amplitude while the arrival times 
and the position along the fibers 
are obtained from the time differences. Cells close in time and space are grouped 
into energy clusters. The cluster energy $E$ is the sum of the cell energies. 
The cluster time $t$ and position $\vec{r}$ are energy-weighted averages. Energy and time 
resolutions are $\sigma_E/E = 0.057/\sqrt{E \mbox{\ (GeV)}}$ and 
$\sigma_t = 57$ ps/$\sqrt{E \mbox{\ (GeV)}} \oplus100$ ps, respectively. 
The cluster space resolution is $\sigma_{\parallel} = 1.4$ cm/$\sqrt{E \mbox{\ (GeV)}}$
along the fibers and $\sigma_{\perp} = 1.3$ cm in the orthogonal direction.

The trigger~\cite{K:trigger} uses both calorimeter and chamber information.
For this analysis the events are selected by the calorimeter trigger, requiring
two energy deposits with $E > 50$ MeV in the barrel or $E > 150$ MeV in
the end-caps. A higher-level cosmic-ray veto rejects events with at least two 
energy deposits above 30 MeV in the outermost calorimeter layer. Data are then analyzed by
an event classification filter~\cite{K:FILFO}, which selects and streams various categories
of events in different output files.

\section{Data sample and event preselection} \label{s:DATASAMPLE}

The data were collected at $\sqrt{s}$ = 1000.1 MeV with electron and positron beams colliding at a small angle with an average transverse momentum of 12.7 MeV in the horizontal plane. The average instantaneous luminosity was $7\times
10^{31}$ cm$^{-2}$s$^{-1}$ and the analysis is based on an integrated luminosity of  242.5 pb$^{-1}$ measured with a precision of 0.3\% recording large angle Bhabha scattering events~\cite{K:Lumi}.

Data are selected with a background rejection filter~\cite{K:FILFO} before event reconstruction. 
A 1/20 sample of unfiltered data, corresponding to about 11
pb$^{-1}$, is also reconstructed to define the preselection
filter used for the analysis and to evaluate its efficiency for event selection. 
The preselection filter requires
\begin{itemize}
 \item at least two energy clusters, neutral (not associated
 to any track) and prompt (with $|t - r/c| < 5\sigma_{t}$);
 \item all prompt neutral clusters are required to have energy $E_{\gamma} > 15$ MeV and polar angle
 $20^{\circ} < \theta_{\gamma} < 160^{\circ}$;
 \item at least one prompt neutral cluster with energy greater than 50 MeV; 
 \item a ratio of the two highest energy neutral prompt clusters to the total
 calorimeter energy $R = (E_{\gamma 1} + E_{\gamma 2})/E_{tot}
 > 0.3$;
 \item 100 MeV $< E_{tot} <$ 900 MeV, to reject low energy
 background events and the high rate processes $e^{+} e^{-} \to
 e^{+} e^{-} (\gamma)$, $e^{+} e^{-} \to \gamma \gamma$.
\end{itemize}

\section{Cross section for $e^+ e^- \to e^+ e^- \eta$ with $\eta \to \pi^+ \pi^- \pi^0$} \label{s:ETA CHARGED}

\subsection{Event selection} \label{s:SELECTION CHARGED}
 
In addition to the preselection, candidate decays $\eta \to \pi^+ \pi^- \pi^0$ should fulfill the following requirements
\begin{itemize}

\item two and only two neutral prompt clusters with $|t - r/c| < 3\sigma_t$ and polar angle $23^{\circ} < \theta_{\gamma} < 157^{\circ}$;
\item at least two tracks with opposite curvature that are extrapolated inside a cylinder $\rho = \sqrt{x^2 + y^2} < 8$ cm and $|z| < 8$ cm centered around the average beam collision point;
\item the distance of the first DC hit to the average beam collision point to be less than 50
cm for both tracks (in case of two or more tracks with the same curvature, the track with best quality
parameters is chosen);
\item sum of the two tracks momenta $|\vec{p}_1| + |\vec{p}_2| < 700$ MeV.
\end{itemize}
To minimize any selection bias and to optimize the selection
efficiency, there is no requirement for the tracks to be associated to clusters in the calorimeter nor that they form
a vertex. The number of selected events is $3.9\times10^6$. A small fraction of fully neutral final states can
survive the two tracks requirement in case of photon conversion $\gamma N \to e^+ e^- N$ or $\pi^0$ Dalitz decay.

\begin{figure}[tbp]
\includegraphics[width = 7.5cm,height=7.5cm]{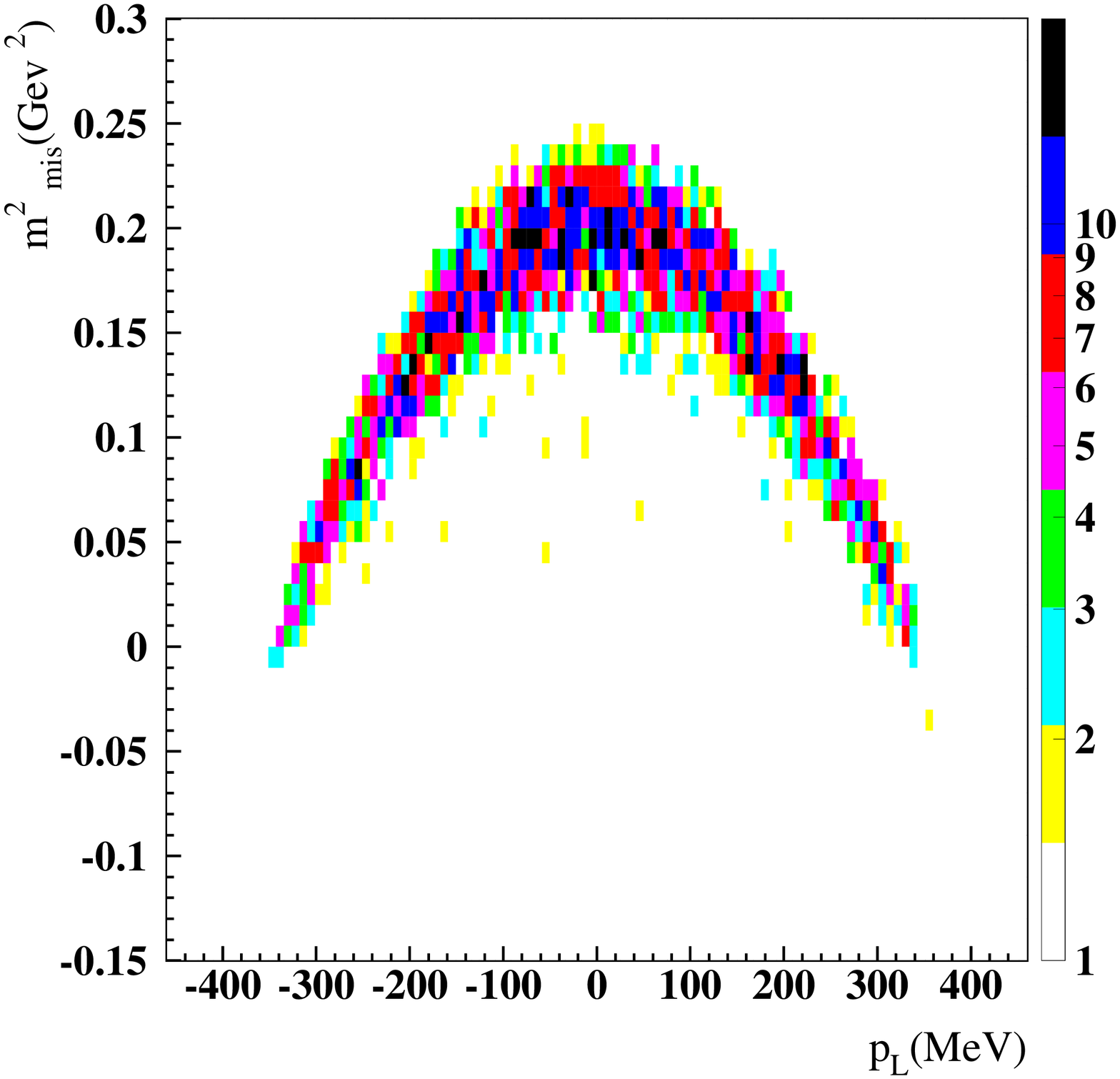}
\hfill
\includegraphics[width = 7.5cm,height=7.5cm]{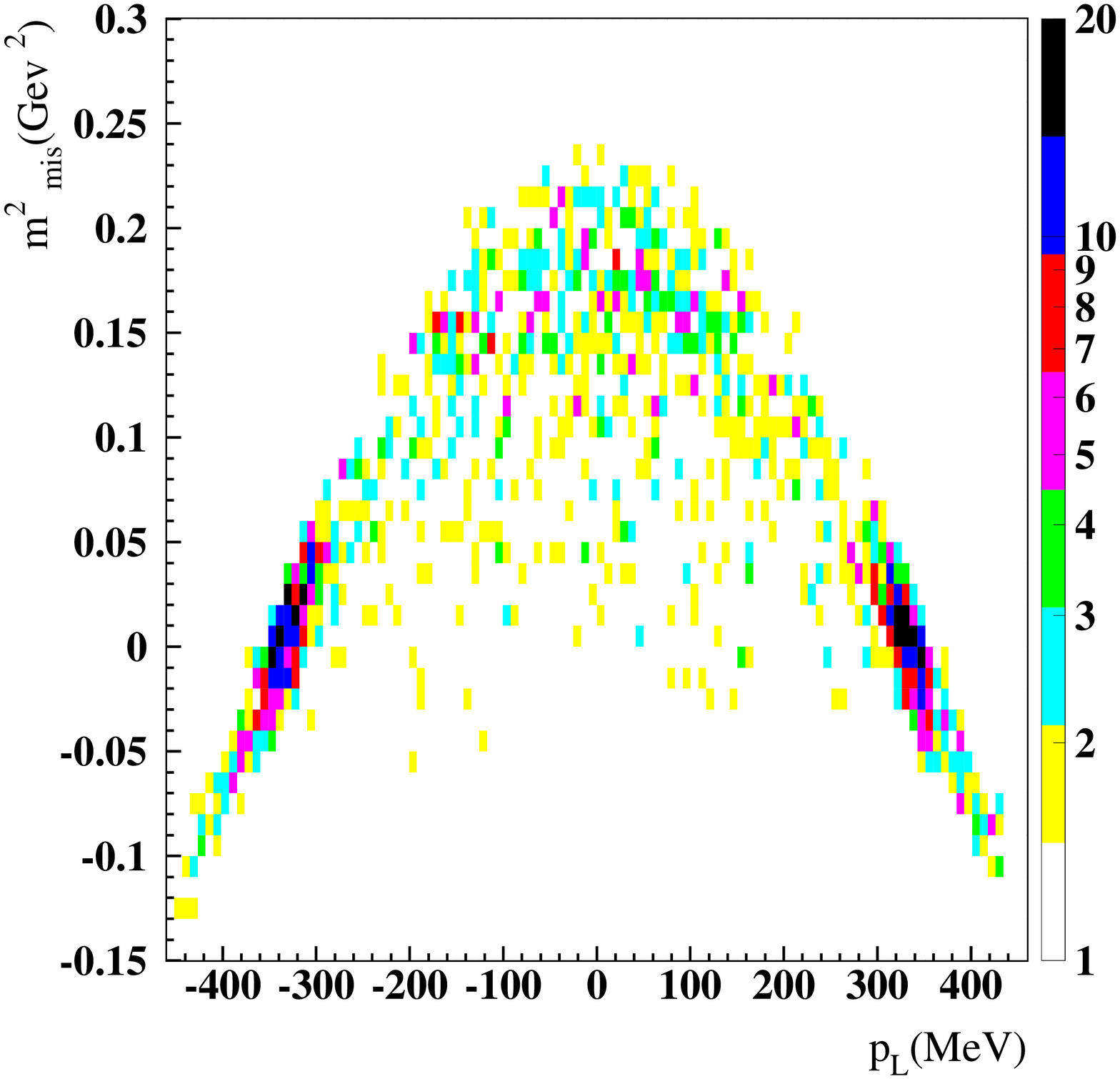}
\caption{Correlation between the $\eta$ longitudinal momentum, $p_{L\eta}$,
 and the squared missing mass, $m^{2}_{mis}$ for the reconstructed events 
that pass the selection cuts of the analysis, for MC signal events (left) and data (right). 
The $e^+e^-\to\eta\gamma$ events, when the monochromatic photon escapes detection, are clearly visible in the data.}
 \label{fig:correl}
\end{figure}

Many background contributions have been considered. 
The $e^+e^- \to \eta \gamma$ process is a source of irreducible background when $\eta$ decays to 
$\pi^+ \pi^- \pi^0$ and the monochromatic photon, $E_{\gamma} = 350$ MeV, is emitted at small
polar angles and is not detected. However, the correlation of the squared missing mass, $m^2_{mis}$ and the $\eta$ longitudinal momentum $p_{L \eta}$ can be used to separate the signal from the background. For the background $p_{L \eta} = E_{\gamma} \cos \theta \simeq 350$ MeV and $m^2_{mis} \simeq 0$ while the signal, for small values of $p_{T \eta}$, is characterized by
$m^2_{mis} \simeq (\sqrt{s} - m_{\eta})^2 + (\sqrt{s}/m_{\eta})\ p^2_{L \eta}$, as shown in Figure~\ref{fig:correl}.

The process $e^+e^- \to \omega \pi^0$, with $\omega \to \pi^+ \pi^- \pi^0$, has four photons in the final state and therefore produces the same final state as the signal when two photons are not detected. The cross section has been measured with data from the same run \cite{K:omegapi0}, $\sigma(e^+e^- \to \omega \pi^0 \to  \pi^+ \pi^- \pi^0 \pi^0) = (5.72 \pm 0.05)$ nb.
The $e^+e^- \to K_L K_S$ events can mimic the signal either when the $K_L$ decays to $\pi^{\pm} \ell^{\mp} \nu$
close to the collision point and $K_S$ decays to $\pi^0 \pi^0$, or when the $K_L$ escapes detection and $K_S \to\pi^0 \pi^0$ is followed by photon conversion in an $e^+ e^-$ pair or by a $\pi^0$ Dalitz decay.
The $e^+e^- \to K^+ K^-$ events can mimic the signal when both kaons decay close to the collision point, either $K^{\pm} \to \pi^{\pm} \pi^0$ or $K^{\pm} \to \pi^{0} \ell^{\pm} \nu$ in coincidence with $K^{\mp} \to \mu^{\mp} \nu$.
Also Bhabha radiative events, $e^+e^- \to e^+e^- \gamma$, given the large cross section, can be a source of background in case of accidental or split clusters.

\subsection{Reconstruction of $\eta \to \pi^+ \pi^- \pi^0$ decay} \label{s:RECO CHARGED}

\begin{figure}[htb]
\centering
\includegraphics[width=8cm,height=8.5cm]{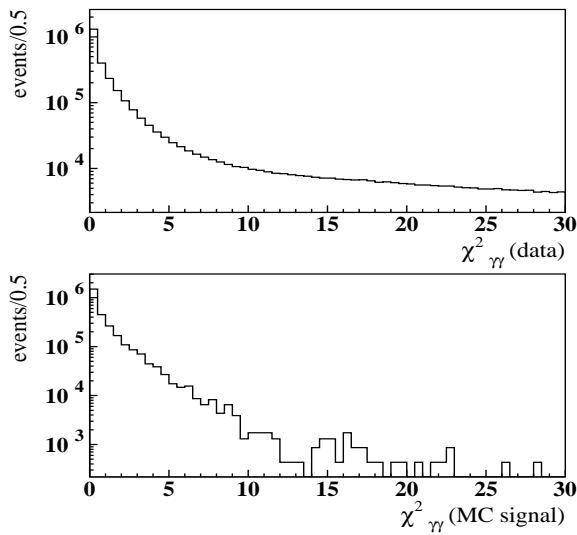}
\caption{Distribution of $\chi^2_{\gamma\gamma}$ 
for data (top) and MC signal events (bottom).}
\label{f:CHI2GG CHARGED}
\end{figure}

To identify the $\pi^0$ meson, clusters are paired choosing the combination that minimizes the difference between the two-cluster invariant mass and the $\pi^0$ mass. This is performed using a pseudo-$\chi^2$ variable
\begin{equation}
\chi^2_{\gamma \gamma} = \frac{(m_{\gamma \gamma} - m_{\pi^0})^2}{\sigma^2_m} 
\qquad \mbox{with} \qquad \sigma_{m} = \frac{m_{\gamma \gamma}}{2} 
\left( \frac{\sigma_{E\gamma i}}{E_{\gamma i}} + \frac{\sigma_{E\gamma j}}{E_{\gamma j}} \right)\,.
\label{eq:CHI2GG}
\end{equation}
The energy resolution function is given in Section~\ref{s:DETECTOR}, the $\gamma\gamma$ invariant mass resolution is dominated by the calorimeter energy resolution while the angle measurement gives a negligible contribution. Figure~\ref{f:CHI2GG CHARGED} shows the distribution of the
$\chi^2_{\gamma \gamma}$ variable for MC signal events and for data. In the following analysis we select events with $\chi^2_{\gamma \gamma} < 8$.

\begin{figure}[tbp]
 \centering
\includegraphics[width=8cm,height=8.5cm]{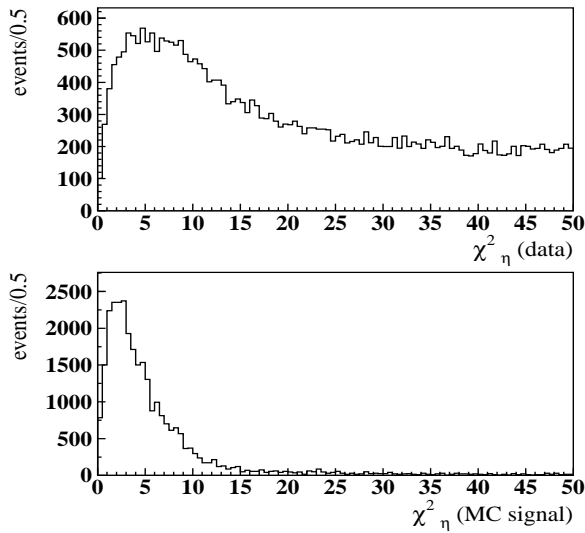}
\caption{Distribution of the $\chi^2_{\eta}$ of the kinematic fit
for data (top) and MC signal events (bottom).}
\label{f:CHI2ETA CHARGED}
\end{figure}

The two tracks momenta are combined with the $\pi^0$ to identify $\eta \to \pi^+ \pi^- \pi^0$ decay candidates, assigning the charged pion mass to the tracks.
A kinematic fit is done requiring the invariant mass of $\pi \pi \gamma \gamma$ equal to the $\eta$ mass. In the fit the energies, $E_i$, the times,  $t_i$, and the coordinates of the cluster centroid position $x_i, y_i, z_i$, for the two clusters are varied. The track momenta are not varied in the minimization since they are measured with
much better precision than the cluster energies. There are four constraints: the promptness of the two clusters, $t_i - r_i/c = 0$, and the mass values $m_{\gamma \gamma} = m_{\pi^0}$, $m_{\pi \pi \gamma \gamma} = m_{\eta}$.
Figure~\ref{f:CHI2ETA CHARGED} shows the distribution of $\chi^2_{\eta}$ from the kinematic fit for MC signal events and for data. We require $\chi^2_{\eta} < 20$
to reduce the $\eta(\to \pi^+ \pi^- \pi^0) \gamma$ background. This process has a long tail in the $\chi^2_{\eta}$ distribution due to events with the monochromatic photon in the detector acceptance and one photon from the $\pi^0$ decay undetected, that are not rejected by the $\chi^2_{\gamma \gamma} < 8$ requirement.

At this stage of the selection, radiative Bhabha scattering, $e^+ e^- \to e^+ e^- \gamma$, and $e^+ e^- \to \gamma \gamma$ annihilation followed by photon conversion are still a source of background. Separation of charged pion from electron/positron tracks is done using a likelihood method when a cluster is associated to the track~\cite{K:EPI LIKELIHOOD}. A cluster is associated if the distance between the centroid and the extrapolation of the track to the calorimeter wall is less than 50 cm. The $e$-$\pi$ likelihood is based on three variables: \textit{i)} the difference of time of flight,  \textit{ii)} the energy of the cluster;  \textit{iii)} the fractions of energy deposited in the first and in the fifth calorimeter layers. 
In this analysis events with a cluster associated to each track and a value of the likelihood estimator $\log \mathcal{L}_{\pi} / \mathcal{L}_e < 0$ for both clusters are rejected.
The background from $e^+ e^- \to \gamma \gamma$ annihilation is reduced requiring that the most energetic cluster satisfies the conditions $E_{\gamma 1} < 230$ MeV and $27.5^{\circ} < \theta_{\gamma 1} < 152.5^{\circ}$.

Opposite curvature track pairs can originate from split tracks. This is due to the track finding algorithm that looks for secondary vertices of kaon decays. Background of split tracks is reduced applying a topological cut based on the correlation between the tracks opening angle, $\alpha_{\pi \pi}$, and the distance between the first DC hits associated to the two tracks by the tracking algorithm. 
The background from kaon decays in $e^+ e^- \to K_L K_S$ and $e^+ e^- \to K^+ K^-$ is reduced applying a cut on the two tracks opening angle $\alpha_{\pi \pi} > 50^{\circ}$. 

Kaon decays are characterized by non prompt energy clusters, thus both the time and the energy assigned by the kinematical fit to the neutral clusters are modified by the fit constraints. This effect is observed in the time and energy pulls built with the two neutral clusters
\begin{equation}
\chi^2_t = \sum_{i \in 2\gamma} \frac{(t^{fit}_i - t^{meas}_i)^2}{\sigma^2_t}\,, \qquad \qquad
\chi^2_E = \sum_{i \in 2\gamma} \frac{(E^{fit}_i - E^{meas}_i)^2}{\sigma^2_E}\,,
\label{eq:CHI2 TE}
\end{equation}
where the superscript \textit{meas} and \textit{fit} indicate the values measured and returned by the fit, respectively.
The pulls are required to satisfy $\chi^2_t < 7$ and $\chi^2_E < 8$.

The selection efficiencies are evaluated with the MC simulation and are listed in Table~\ref{t:EFF_CHARGED} for the signal and the most relevant background sources. 
The column Selection includes the efficiency of the trigger, 
the background filter 
and the data filters described in 
Sections~\ref{s:DATASAMPLE} and~\ref{s:SELECTION CHARGED}. The trigger efficiency is controlled by comparison of 
the calorimeter trigger with a complementary trigger based on the drift chamber hit patterns~\cite{K:trigger}. 
A sample of unfiltered data is used to control the filter efficiency. 

The signal is simulated with different values of the $b_{\eta}$ parameter \footnote{The values chosen are $b_{\eta}=$ 0, 0.7, 1.0, 1.5, 1.64, 1.80, 1.94, 2.00, 2.24 GeV$^{-2}$} of the form factor in equation 
(\ref{eq:TRANSFF}) and the fit to derive the signal yield is repeated for each value. The values of efficiencies shown in Table~\ref{t:EFF_CHARGED} correspond to $b_{\eta} = 1.94$ GeV$^{-2}$. 

\begin{table}[tbp]
 \centering
 \begin{tabular}{|c|c|c|}
\hline
 Final state & Selection & Global  \\
                  & efficiency (\%) & efficiency (\%)  \\
 \hline
 $e^+ e^- \eta$ & 34.4 $\pm$ 0.3 & 20.8 $\pm$ 0.3 \\
 \hline
 $\eta (\to \pi^+ \pi^- \pi^0)\gamma$ & 13.3 & 1.93 \\
 $\eta (\to \pi^+ \pi^- \gamma)\gamma$ & 43.9 & 0.0090 \\
  $\eta (\to \mbox{neutral})\gamma$ & 0.185 & 0.00030 \\
 $\omega \pi^0$ & 3.08 & 0.023 \\
 $K_L K_S$ & 0.169 & 0.0059 \\
 $K^+ K^-$ & 0.423 & 0.0075 \\
 $e^+ e^- \gamma$ & 0.447 & $<$ 0.0004 \\
 \hline
  \end{tabular}
 \caption{Selection efficiency, in \%, for the signal and the most relevant backgrounds. The column Selection includes the efficiency of the trigger, the background filter 
 and the data filters described in Sections~\ref{s:DATASAMPLE} and~\ref{s:SELECTION CHARGED}. The global efficiency for $e^+e^-\gamma$ is derived before the cut on the $e$-$\pi$ likelihood.}
 \label{t:EFF_CHARGED}
\end{table}

\subsection{Cross section evaluation} \label{s: XSECTION CHARGED}

The analysis cuts described in Section~\ref{s:RECO CHARGED} select 2977 events. The number of signal events is derived with a 2-dimensional fit to the data. The variables used to discriminate the signal from background are the squared missing mass and the $\eta$ transverse momentum in the interval $-0.15$ GeV$^2 < m^2_{mis} < 0.25$ GeV$^2$ and $p_{T \eta} <$ 300 MeV that contains 2720 events. The fit to the data is done using the simulated shapes for the signal and backgrounds, the weights are left free except for $\eta(\to \pi^+ \pi^- \pi^0) \gamma$ whose cross section and error, measured in the same data sample (see Section~\ref{s:ETAG XSECTION}), is used as a constraint in the fit. The fit returns the fraction of data events $f_i = n_i/n_{tot}$ with the constraint $\sum_i f_i = 1$.

\begin{figure}[tbp]
 \includegraphics[width=8cm,height=8cm]{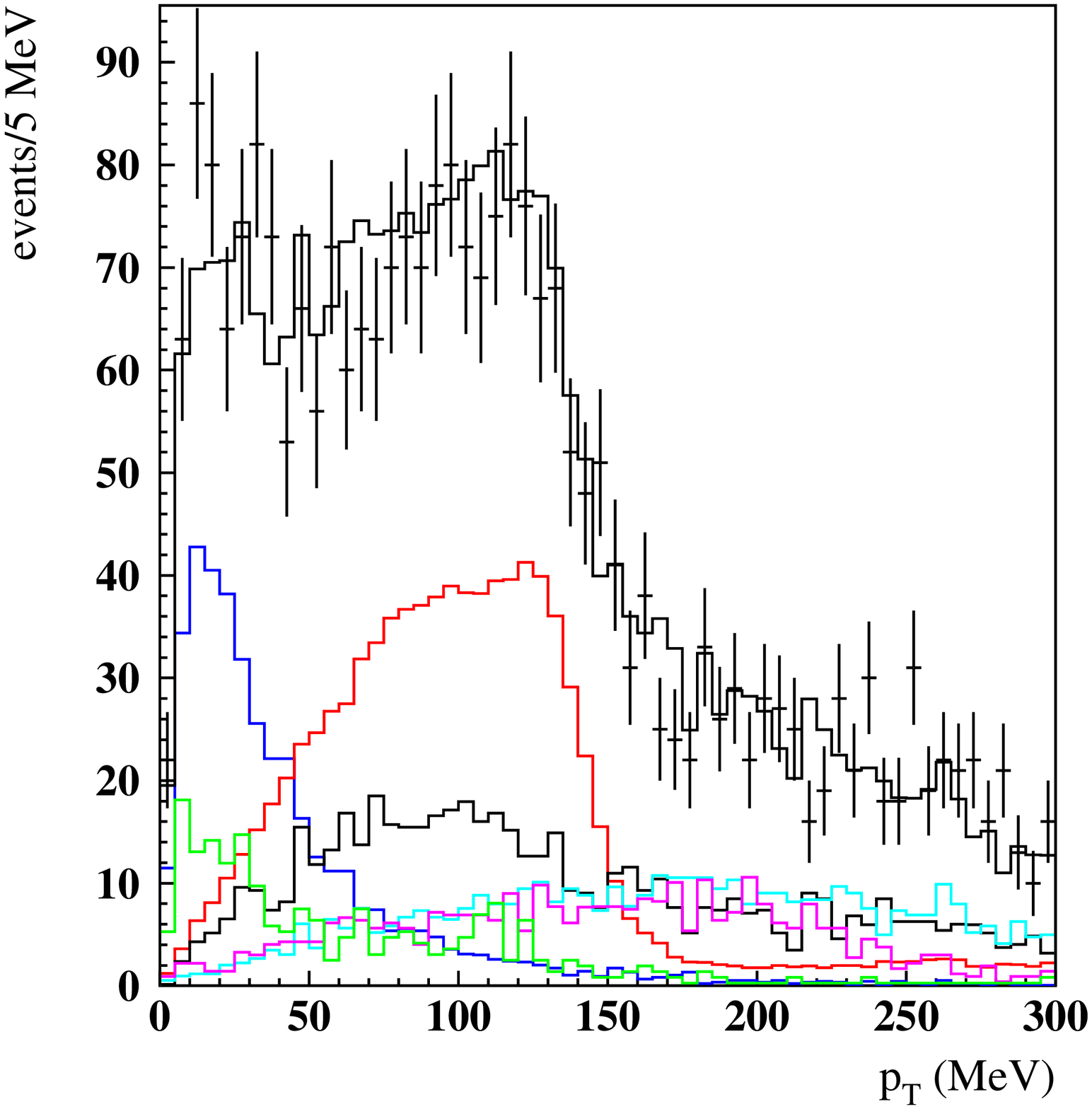}
\hspace{-0.5cm}
 \includegraphics[width=8cm,height=8cm]{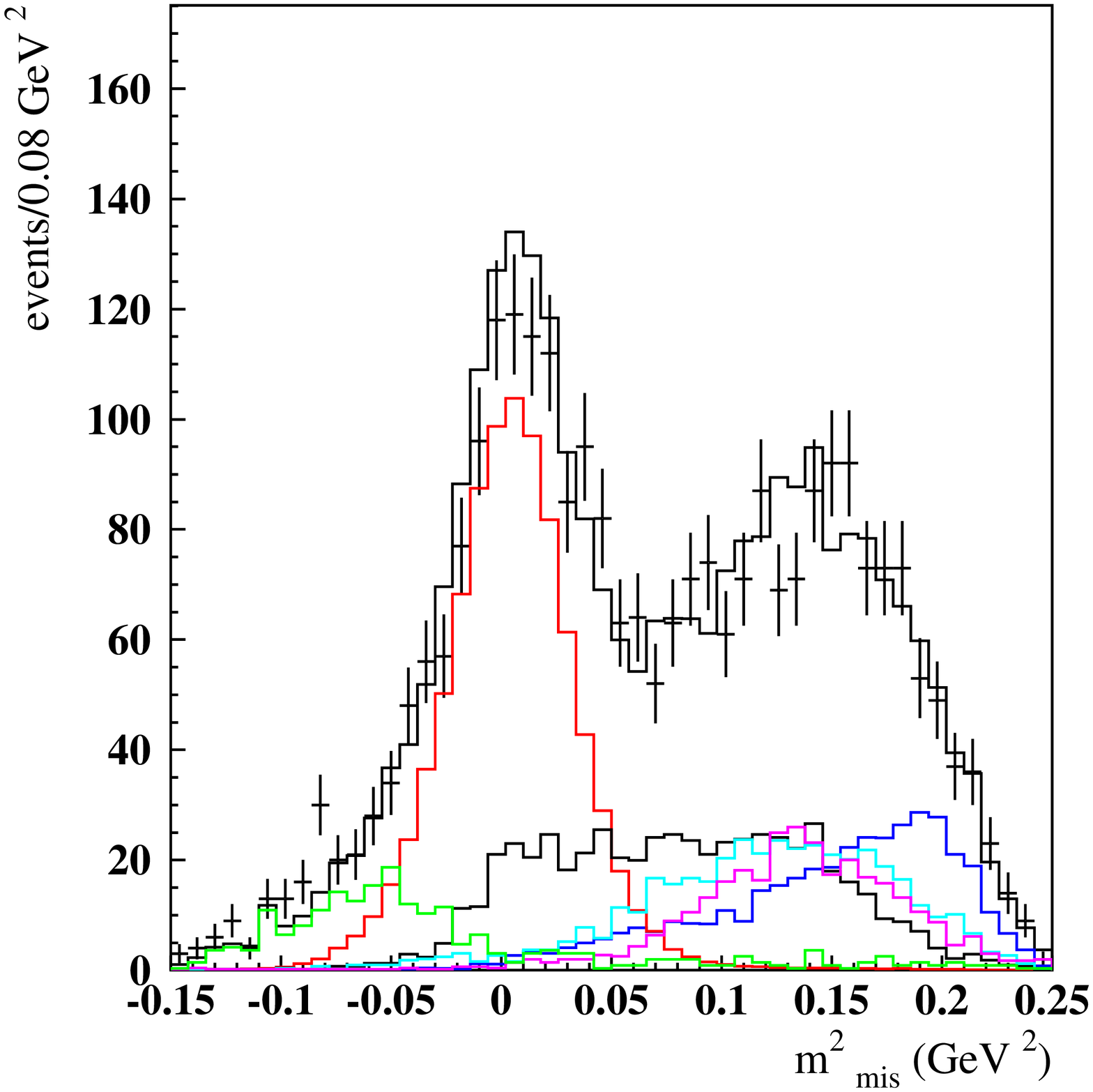}
 \caption{Projections of the 2-dimensional fit. Left: distribution of the transverse momentum of the $\pi^+\pi^-\gamma\gamma$ system. 
Right: distribution of the squared missing mass. 
The contribution of the signal is blue, $e^+e^-\to\eta\gamma$ is red, $e^+e^-\to\omega\pi^0$ is black, 
$e^+e^-\to e^+e^-\gamma$ is green, $e^+e^-\to K^+K^-$ is light blue and $e^+e^-\to K_SK_L$ is purple.}
\label{f:MMISS2vsPT CHARGED}
\end{figure}

\begin{figure}[tbp]
 \centering
 \includegraphics[width = 8cm,height=8cm]{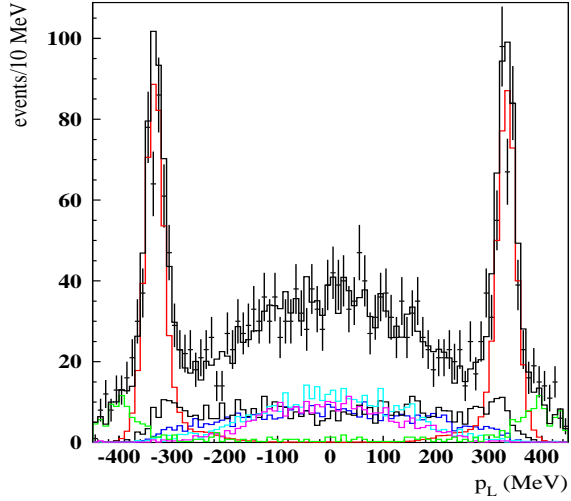}
 \caption{Distribution of the longitudinal $\pi^+\pi^-\gamma\gamma$ momentum. 
The contribution of the signal is blue, $e^+e^-\to\eta\gamma$ is red, $e^+e^-\to\omega\pi^0$ is black, 
$e^+e^-\to e^+e^-\gamma$ is green, $e^+e^-\to K^+K^-$ is light blue and $e^+e^-\to K_SK_L$ is purple.}
 \label{f:PL CHARGED}
\end{figure}

The projections of the $m^2_{mis}\times p_{T \eta}$ distribution are shown in Figure~\ref{f:MMISS2vsPT CHARGED} for the data and the backgrounds weighted by their fractions $f_i$, and the $p_{L \eta}$ distribution is shown in Figure~\ref{f:PL CHARGED}. The most relevant background is $e^+ e^- \to \eta \gamma$ characterized by $m^2_{mis} \simeq 0$ and $p_{L \eta} \simeq \pm 350$ MeV. Table~\ref{t:FIT_CHARGED} lists the fraction of events returned by the fit using the signal efficiency evaluated with $b_{\eta} = 1.94$ GeV$^{-2}$, the fit is repeated for all the other $b_{\eta}$ values used in evaluating the efficiency. The distributions of the variables used in the event selection are compared for data and MC simulation, weighted by the fractions $f_i$ returned by the fit, and good agreement is observed. The fit finds $394 \pm 29$ signal events. 

\begin{table}[tbp]
 \centering
 \begin{tabular}{|c|c|}
\hline
 Final state & Fraction of events (\%)  \\
 \hline
 $e^+ e^- \eta$ & 14.49 $\pm$ 1.06  \\
 \hline
 $\eta \gamma$ & 32.02 $\pm$ 0.54 \\
 $\omega \pi^0$ &  20.48 $\pm$ 1.81 \\
 $K_L K_S$ &  11.36 $\pm$ 1.70 \\
 $K^+ K^-$ & 15.13 $\pm$ 1.81 \\
 $e^+ e^- \gamma$ & 7.54 $\pm$ 0.87 \\
 \hline
  \end{tabular}
 \caption{Fraction of events, in \%, for the signal and the most relevant backgrounds.}
 \label{t:FIT_CHARGED}
\end{table}

The contributions to the systematic error are evaluated by varying the analysis cuts by the r.m.s. width of the distributions of each variable: $\chi^2_{\gamma \gamma}$, $\chi^2_{\eta}$, $E_{\gamma 1}$, $\theta_{\gamma 1}$, $\alpha_{\pi \pi}$, $\chi^2_t$, $\chi^2_E$, accounting for their correlation. 
This results in a systematic relative error of $-2.4\%$ to $+2.6\%$. The contributions are listed in Table~\ref{t:Table66}.

\begin{table}[tbp]
\vspace{0.5cm}
\centering
 \begin{tabular}{|c|c|c|}
\hline
\ Variable & Range & $\delta\sigma/\sigma (\%)$ \\
\hline
 $\chi^2_{\gamma\gamma}$ & 6.6 - 10.8 & +0.67 \quad -0.73 \\ 
\hline
 $\chi^2_{\eta}$ &  18.5 - 23.5 & +0.06 \quad -0.68 \\ 
\hline
 $E_{\gamma 1}$ & 210 MeV - 250 MeV & -1.17 \quad -0.33 \\ 
\hline
 $\theta_{\gamma 1}$ & 26.5$^\circ$/153.5$^\circ$ - 28.5$^\circ$/151.5$^\circ$ & +1.21 \quad +0.46 \\
\hline
 $\chi^2_{\rm t}$ & 6 - 8 & +1.10 \quad -1.22 \\ 
\hline
 $\chi^2_{\rm E}$ & 7 - 9 & +1.89 \quad -1.39 \\ 
\hline
$\alpha_{\pi^+\pi^-}$ & 48$^\circ$ - 52$^\circ$ & -0.21 \quad +0.20 \\ 
\hline
 \end{tabular}
 \caption{Systematic errors determined varying the cuts for each variable for the $\sigma(e^{+} e^{-} \to e^{+} e^{-} \eta \to e^+e^- \pi^+\pi^-\pi^0)$ measurement.}
\label{t:Table66}
\end{table}
 
The MC simulation statistical error of 1.4\% (Table~\ref{t:EFF_CHARGED}) is added in quadrature, 
the uncertainties in the form factor and in the branching ratio are kept separate to
account for correlations between the two $\eta$ decay modes. 
The change of the result due to the variation of $b_{\eta}$ 
in the transition form factor formula leads to a $2.0\%$ fractional error. 
We obtain 
$\sigma(e^+ e^- \to e^+ e^- \eta \to e^+ e^- \pi^+ \pi^- \pi^0) = (7.84 \pm 0.57_{\mbox{stat}} 
\pm 0.23_{\mbox{syst}} \pm 0.16_{\mbox{\footnotesize{FF}}})$ pb. 
Using for the branching fraction the value $BR(\eta \to \pi^+ \pi^- \pi^0) = 0.2274 \pm 0.0028$~\cite{PDG}, 
we obtain 
\begin{equation}
\sigma(e^+ e^- \to e^+ e^- \eta) = (34.5 \pm 2.5_{\mbox{stat}} \pm 1.0_{\mbox{syst}} \pm 0.7_{\mbox{\footnotesize{FF}}} 
\pm 0.4_{\mbox{\footnotesize{BR}}}) \mbox{\ pb}\,.
\label{eq:SIGMA CHARGED}
\end{equation}

\section{Cross section for $e^+ e^- \to e^+ e^- \eta$ with $\eta \to \pi^0 \pi^0 \pi^0$} \label{s:ETA NEUTRAL}

\subsection{Event selection} \label{s:SELECTION NEUTRAL}

In addition to the preselection described in Section~\ref{s:DATASAMPLE}, candidate decays $\eta \to 3\pi^0$ should fulfill the following requirements
\begin{itemize}

\item six and only six neutral prompt clusters with $E_{\gamma} > 15$ MeV, $|t - r/c| < 3\sigma_t$ and polar angle $23^{\circ} < \theta_{\gamma} < 157^{\circ}$;
\item no tracks in the drift chamber.
\end{itemize}

The number of selected events is 9857.
Many background contributions have been considered. As in the charged decay analysis, the $e^+e^- \to \eta \gamma$ process is a source of irreducible background when $\eta$ decays to $3 \pi^0$ and the recoil photon is not detected.
The process $e^+ e^- \to \omega \pi^0$ with $\omega \to \pi^0 \gamma$ produces 5 photons in the final state and is important in case of accidental or split clusters. The cross section has been measured with the same data set~\cite{K:omegapi0}: 
$\sigma(e^+ e^- \to \omega\pi^0\to\pi^0 \pi^0 \gamma) = (0.550 \pm 0.005)$ nb. The process $e^+ e^- \to a_0(980) \gamma \to \eta \pi^0 \gamma$ can mimic the signal when $\eta$ decays to $3 \pi^0$ and three photons are not detected, or it decays to $\gamma \gamma$ with split or accidental clusters. Similarly for $e^+ e^- \to f_0(980) \gamma \to \pi^0 \pi^0 \gamma$ and $e^+ e^- \to \eta' \gamma$ when $\eta'$ decays to neutrals. Also the process $e^+ e^- \to K_L K_S$ with $K_S \to \pi^0 \pi^0$ and undetected $K_L$ can mimic the signal in case of split or accidental clusters.

\subsection{Reconstruction of $\eta \to 3 \pi^0$ decay} \label{s:RECO NEUTRAL}

\begin{figure}[tbp]
\centering
\includegraphics[width=8cm,height=8.5cm]{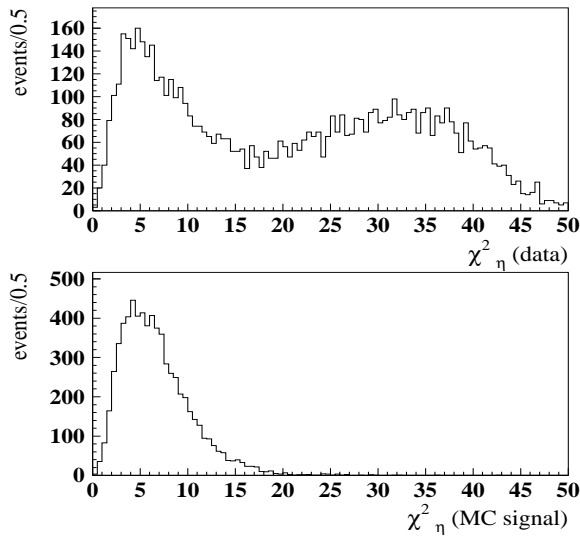}
\caption{Distribution of $\chi^2_{\eta}$ for data (top) and MC signal events (bottom).}
\label{f:CHI2 6G}
\end{figure}

The six photons are paired choosing the combination that minimizes the difference between the $\gamma \gamma$ invariant mass of the pairs and the mass of the $\pi^0$ as described in Section~\ref{s:RECO CHARGED}. In the following analysis we select events with $\chi^2_{6\gamma} < 14$.
A kinematic fit is done requiring the $6 \gamma$ invariant mass to be equal to the $\eta$ mass. In the fit the energies, $E_i$, the times,  $t_i$, and the coordinates of the centroid positions $x_i, y_i, z_i$, for the six clusters are varied. 
There are seven constraints: the promptness of the six clusters $t_i - r_i/c = 0$ and $m_{6 \gamma} = m_{\eta}$. Figure~\ref{f:CHI2 6G} shows the distribution of $\chi^2_{\eta}$ from the kinematic fit for MC signal events and for data, we require $\chi^2_{\eta} < 20$ to reduce the background. 

MC simulation shows that $e^+ e^- \to \eta \gamma$ gives a large contribution to the tail of the distribution when 
the monochromatic photon is in the acceptance and is wrongly paired with a photon from $\eta$ decay. 
In this case it also produces an enhancement at large values 
of the $6 \gamma$ invariant mass distribution. 
To reduce the background we require the highest energy neutral cluster to have $E_{\gamma 1} < 260$ MeV and the six-photon invariant mass $m_{6\gamma} < 630$ MeV.

The selection efficiencies are evaluated with the MC simulation described in Section~\ref{s:MODEL} and are listed in Table~\ref{t:EFF_NEUTRAL} for the signal and the most relevant background sources. 
\begin{table}[tbp]
 \centering
 \begin{tabular}{|c|c|c|}
\hline
 Final state & Selection & Global  \\
                  & efficiency (\%) & efficiency (\%)  \\
 \hline
 $e^+ e^- \eta$ & 30.9 $\pm$ 0.3 & 28.6 $\pm$ 0.3 \\
 \hline
 $\eta (\to \pi^0 \pi^0 \pi^0)\gamma$ & 10.9 & 2.14 \\
 $\omega \pi^0$ & 0.145 & 0.0077 \\
 $K_L K_S$ & 0.0126 & 0.0073 \\
 $a_0(980) \gamma$ & 2.70 & 0.85 \\
 $f_0(980) \gamma$ & 0.147 & 0.0070 \\
 $\eta' \gamma$ & 2.13 & 0.212 \\
 \hline
  \end{tabular}
 \caption{Selection efficiency, in \%, for the signal and the most relevant backgrounds. The column Selection includes the efficiency of the trigger, the background filter  
and the data filters described in Sections~\ref{s:DATASAMPLE} and~\ref{s:SELECTION NEUTRAL}.}
 \label{t:EFF_NEUTRAL}
\end{table}

\subsection{Cross section evaluation} \label{s: XSECTION NEUTRAL}

The number of signal events is derived with a 2-dimensional fit to the data. 
The distributions used to discriminate the signal from background are 
the squared missing mass and the $\eta$ longitudinal momentum in the interval 
-0.15 GeV$^2 < m^2_{mis} <$ 0.35 GeV$^2$ and -450 MeV $ < p_{L \eta} <$ 450 MeV that contains 2166 events. 
The fit to the data is done using the simulated shapes for the signal and backgrounds and the fit returns the fraction of data events $f_i = n_i/n_{tot}$ with the constraint $\sum_i f_i = 1$. 

The contribution of all backgrounds, except $\eta \gamma$ production, is very small, below the statistical sensitivity of the fit. The contribution of $\omega \pi^0$ derived from the value of the $e^+ e^- \to \omega \pi^0$ cross section~\cite{K:omegapi0} is $f_{\omega \pi^0} = 0.47\%$. The contributions of $a_0(980) \gamma$, $\eta' \gamma$ and $K_L K_S$ expected extrapolating the measurements at the $\phi$ peak are negligible.

\begin{figure}[tbp]
 \includegraphics[width=8cm,height=8cm]{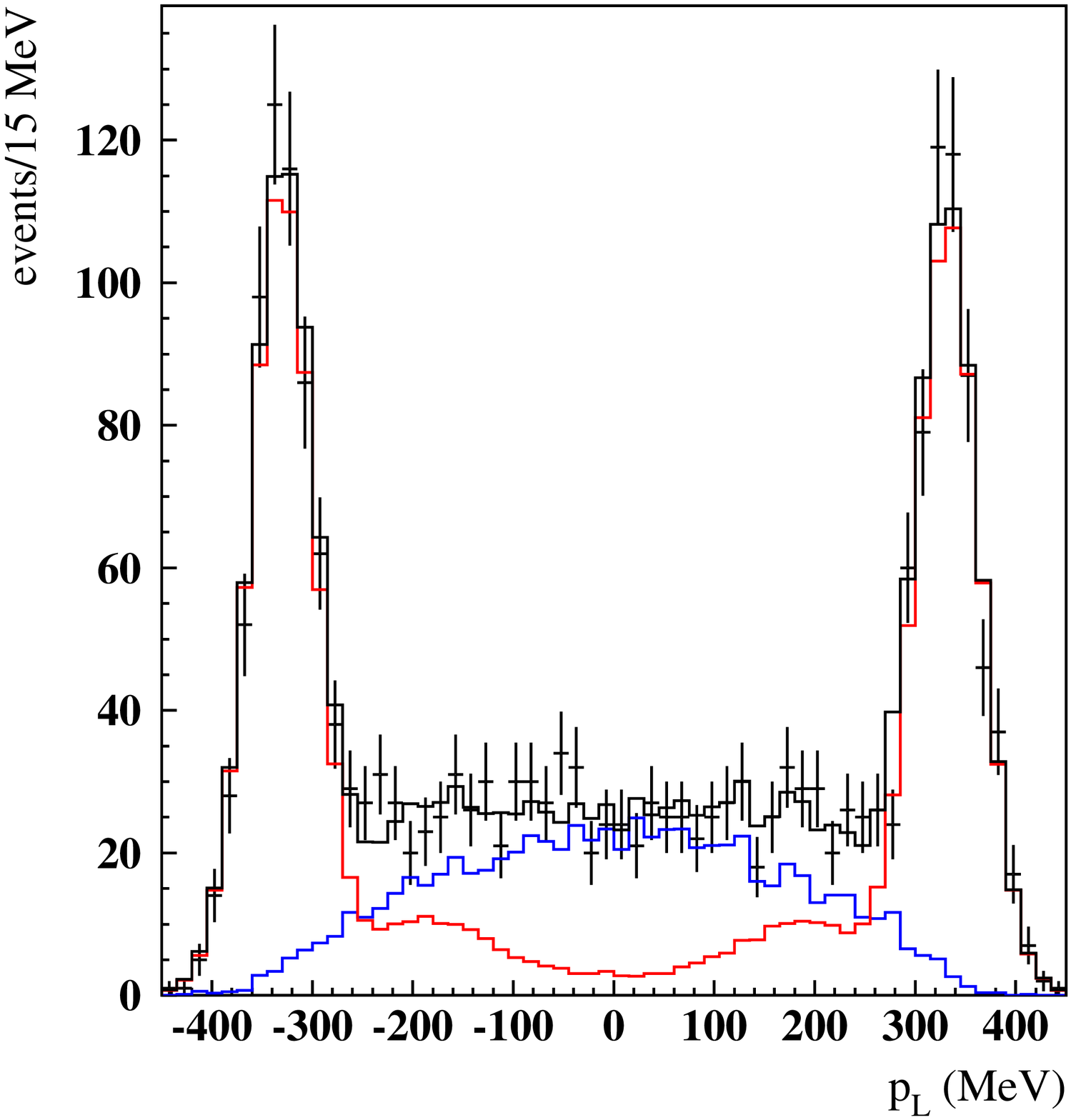}
\hspace{-0.5cm}
 \includegraphics[width=8cm,height=8cm]{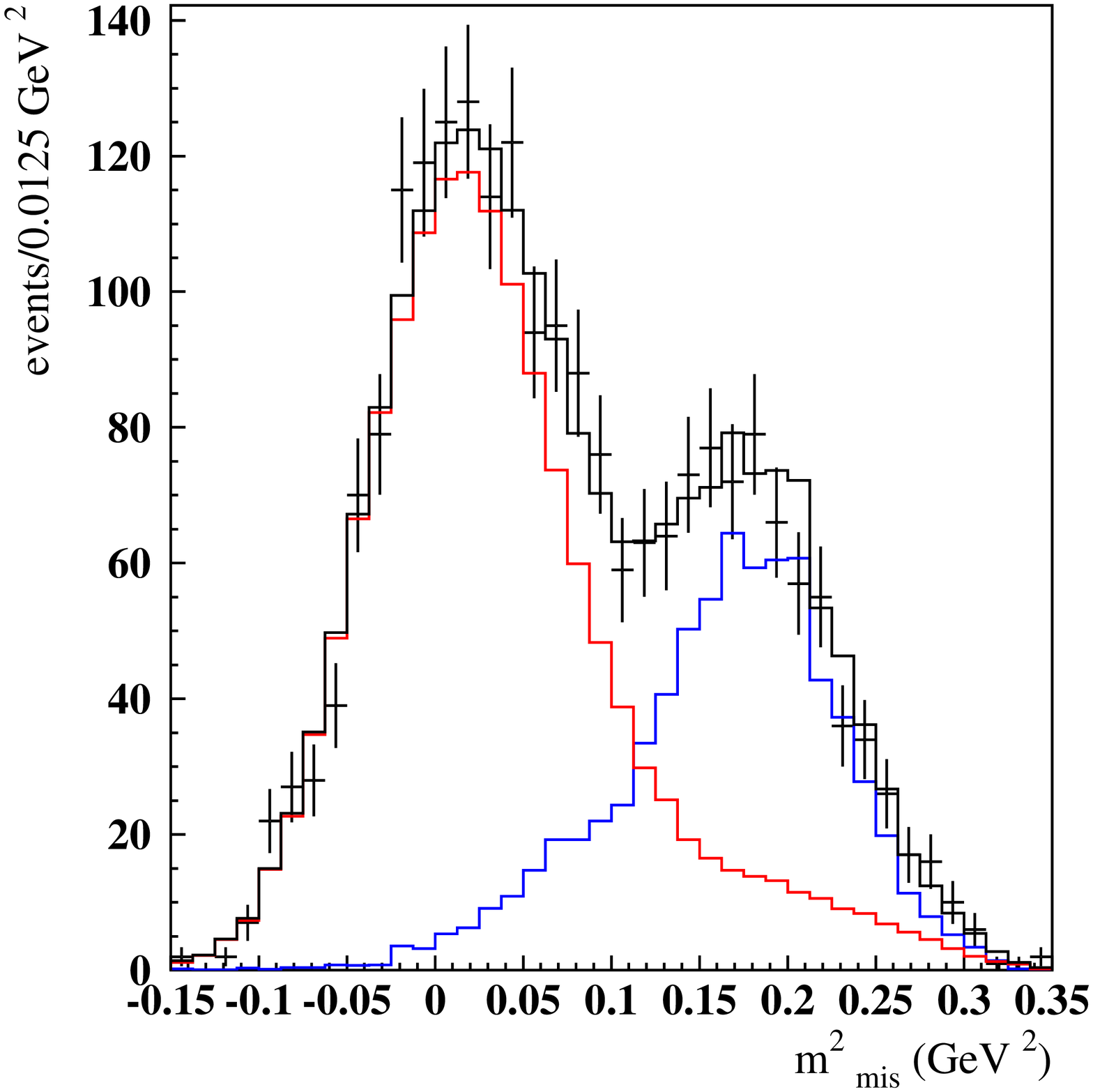}
 \caption{Projections of the 2-dimensional fit. Left: distribution of the $6\gamma$ longitudinal momentum. 
Right: distribution of the squared missing mass. The contribution of the signal is blue, $e^+e^-\to\eta\gamma$ is red.} 
\label{f:MMISS2vsPL NEUTRAL}
\end{figure}

\begin{figure}[tbp]
\centering
 \includegraphics[width=8cm,height=8cm]{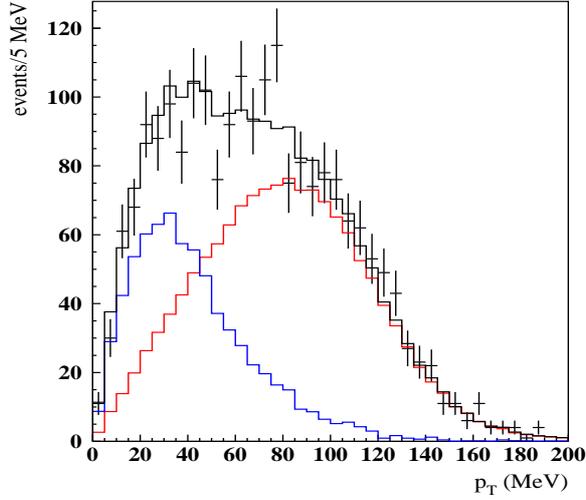}
 \caption{Distribution of the $6\gamma$ transverse momentum. The contribution of the signal is blue, $e^+e^-\to\eta\gamma$ is red} 
\label{f:PT NEUTRAL}
\end{figure}

The fit with two components gives $f_{ee\eta} = (33.4 \pm 1.5)\%$ and $f_{\eta \gamma} = (66.6 \pm 1.9)\%$ using the signal efficiency evaluated with $b_{\eta} = 1.94$ GeV$^{-2}$. The fit is repeated for all the other values. The projections of the $m^2_{mis}\times p_{L \eta}$ distribution are shown in Figure~\ref{f:MMISS2vsPL NEUTRAL} for the data and the background weighted by their relative factors $f_i$, and the $p_{T \eta}$ distribution 
is shown in Figure~\ref{f:PT NEUTRAL}. 

The contributions to the systematic error are evaluated by varying the analysis cuts by the r.m.s. width of the distributions of each variable: $\chi^2_{\gamma \gamma}$, $\chi^2_{\eta}$, 
$E_{\gamma 1}$, $m_{6\gamma}$,  accounting for their correlation. 
This results in a systematic relative error of $-1.5\%$ to $+2.6\%$. The contributions are listed in Table~\ref{t:Table77}.

\begin{table}[tbp]
\centering
 \begin{tabular}{|c|c|c|}
 \hline
\  Variable & Range & $\delta\sigma/\sigma (\%)$ \\
\hline
 $\chi^2_{\gamma\gamma}$ & 12 - 16 & -0.51 \quad +0.83 \\ 
\hline
 $\chi^2_{\eta}$ & 17 - 23 & -0.26 \quad -0.68 \\ 
\hline
 $M_{6\gamma}$ & 610 MeV - 650 MeV & -1.33 \quad +2.38 \\ 
\hline
 \end{tabular}
 \caption{Systematic errors determined varying the cuts for each variable for the $\sigma(e^{+} e^{-} \to e^{+} e^{-} \eta \to e^+e^- 3\pi^0)$ measurement. Varying the cut on $E_{\gamma 1}$ gives a negligible contribution.}
\label{t:Table77}
\end{table}

The
MC simulation statistical error of 1.0\% (Table~\ref{t:EFF_NEUTRAL}) is added in quadrature, 
the errors due to knowledge of the form factor and to the branching ratio are kept separate. 
The changes of the result due to the variation of $b_{\eta}$ 
in the transition form factor formula lead to a $0.7\%$ fractional error. 
We obtain 
$\sigma(e^+ e^- \to e^+ e^- \eta \to e^+ e^- 3 \pi^0) = (10.43 \pm 0.48_{\mbox{stat}} 
\pm 0.29_{\mbox{syst}} \pm 0.07_{\mbox{\footnotesize{FF}}})$ pb. 
The analysis of the systematic uncertainties of the $e^+e^-\to\eta\gamma$ measurement leads to a relative error of $0.6\%$: 
we obtain $\sigma(e^+ e^- \to \eta \gamma \to 3\pi^0 \gamma) = (278.0 \pm 8.1_{\mbox{stat}} \pm 1.7_{\mbox{syst}})$ pb.
Using for the branching fraction the value $BR(\eta \to \pi^0 \pi^0 \pi^0) = 0.3257 \pm 0.0023$~\cite{PDG}, we obtain
\begin{equation}
\sigma(e^+ e^- \to e^+ e^- \eta) = (32.0 \pm 1.5_{\mbox{stat}} \pm 0.9_{\mbox{syst}} \pm 0.2_{\mbox{\footnotesize{FF}}} \pm 0.2_{\mbox{\footnotesize{BR}}}) \mbox{\ pb}
\label{eq:SIGMA NEUTRAL}
\end{equation}
and
\begin{equation}
\sigma(e^+ e^- \to \eta \gamma) = (853 \pm 25_{\mbox{stat}} \pm 5_{\mbox{syst}} 
\pm 6_{\mbox{\footnotesize{BR}}}) \mbox{\ pb}\,.
\label{eq:SIGMA ETAG000}
\end{equation}

\section{Determination of $\Gamma(\eta \to \gamma \gamma)$} \label{s: EEETA XSECTION}

The two values of the cross section in equations~(\ref{eq:SIGMA CHARGED}) and~(\ref{eq:SIGMA NEUTRAL}) are combined accounting for the following sources of correlation:
\begin{itemize}
\item systematic uncertainties are correlated due to the requirements on the neutral prompt clusters, the photon energy, time and position resolutions common to both selections and fit procedures;
\item the determination of the signal efficiencies for the two measurements that share the same transition form factor;
\item the systematic error in the measurement of the luminosity~\cite{K:Lumi};
\item the correlation between the $\eta\to\pi^+\pi^-\pi^0$ and $\eta\to 3\pi^0$ branching ratios~\cite{PDG}.
\end{itemize}
From the combination of the two measurements we derive
\begin{equation}
\sigma(e^+ e^- \to e^+ e^- \eta) = (32.7 \pm 1.3_{\mbox{stat}} \pm 0.7_{\mbox{syst}}) \mbox{\ pb}\,.
\label{eq:SIGMA EEETA}
\end{equation}

The partial width of the $\eta$ meson, $\Gamma(\eta \to \gamma \gamma)$, can be determined from equations~(\ref{eq:eeXSECTION}) and~(\ref{eq:ggXSECTION}). The $\gamma \gamma$ differential luminosity is calculated following reference~\cite{NPP:sigma}, the program computes also the transition form factor as parametrized in equation~(\ref{eq:TRANSFF}), for the same values of the $b_{\eta}$ parameter used in evaluating the $e^+e^-\to e^+e^-\eta$ cross section. 
Since the values of the 4-momenta $q_1$ and $q_2$ sampled in the two decay modes analyzed in Sections~\ref{s:ETA CHARGED} and~\ref{s:ETA NEUTRAL} can be slightly different, the partial width is determined separately for the two decays. 
The theoretical error in evaluating $\sigma(\gamma \gamma \to \eta)$ has been added to the systematic error due to the form factor. From the two values of the $e^+ e^- \to e^+ e^- \eta$ cross section,~(\ref{eq:SIGMA CHARGED}) and~(\ref{eq:SIGMA NEUTRAL}), we derive
\begin{equation}
\begin{split}
\eta \to \pi^+ \pi^- \pi^0 
\qquad 
\Gamma(\eta \to \gamma \gamma) &= (548 \pm 40_{\mbox{stat}} \pm 16_{\mbox{syst}} \pm 14_{\mbox{\footnotesize{FF}}} 
\pm 7_{\mbox{\footnotesize{BR}}}) \mbox{\ eV}\,, \\
\eta \to \pi^0 \pi^0 \pi^0 
\qquad 
\Gamma(\eta \to \gamma \gamma) &= (509 \pm 23_{\mbox{stat}} \pm 14_{\mbox{syst}} \pm 8_{\mbox{\footnotesize{FF}}}
\pm 4_{\mbox{\footnotesize{BR}}}) \mbox{\ eV}\,.
\end{split}
\end{equation}
The two measurements are combined accounting for their correlations to derive
\begin{equation}
\Gamma(\eta \to \gamma \gamma) = (520 \pm 20_{\mbox{stat}} \pm 13_{\mbox{syst}}) \mbox{\ eV}\,.
\label{eq:GAMMA}
\end{equation}

\section{Measurement of the cross section for $e^+ e^- \to \eta \gamma$} \label{s:ETAG XSECTION}

The most relevant background in the measurement of the $e^+ e^- \to e^+ e^- \eta$ cross section is due to the radiative process  $e^+ e^- \to \eta \gamma$. The value of the cross section has been used as a constraint in the fit in case of the $\eta \to \pi^+ \pi^- \pi^0$ decay while it has been derived as a by-product of the analysis of the $\eta \to 3 \pi^0$ decay. The cross section has been measured by the SND experiment~\cite{SND:etagamma} at VEPP-2M in the range $\sqrt{s} = (0.6-1.38)$ GeV, but with less precision than needed to control the analysis of $e^+e^-\to e^+e^-\eta$.

The cross section for $e^+ e^- \to \eta \gamma$ is measured exploiting the $\eta \to \pi^+ \pi^- \pi^0$ decay using the same data sample and the same preselection procedure described in Sections~\ref{s:DATASAMPLE}
and~\ref{s:SELECTION CHARGED} with the only difference that in this case events with three and only three neutral prompt clusters are selected. The event selection aims at finding two tracks of opposite curvature, compatible with being due to $\pi^{\pm}$, two neutral prompt clusters compatible with being originated by a $\pi^0$ decay, and a third neutral prompt cluster compatible with the photon recoiling against the $\pi^+ \pi^- \pi^0$ system. 
 
Several background processes have been considered.
$e^+e^- \to \omega \pi^0$ with $\omega \to \pi^+ \pi^- \pi^0$  is characterized by two tracks and four photons and can simulate the signal if one photon is not detected.
$e^+e^- \to \pi^+\pi^-\pi^0 \gamma$ has the same configuration as the signal. 
$e^+ e^- \to K_L K_S$ can mimic the signal when $K_L$ decays to $\pi^{\pm} \ell^{\mp} \nu$ close to the collision point and $K_S$ decays to $\pi^0 \pi^0$ but one photon is not detected.
$e^+ e^- \to K^+ K^-$ can mimic the signal when both kaons decay close to the collision point to $\pi^{\pm} \pi^0, \pi^{\mp} \pi^0$ and one photon is not detected, or decay to $\pi^{\pm} \pi^0, \mu^{\mp} \nu$ and the additional photon originates from split or accidental clusters.
$e^+ e^- \to \pi^+ \pi^- \pi^0$ and $e^+ e^- \to \pi^+ \pi^- \gamma$ can mimic the signal in case of one or two accidental or split clusters.
$e^+ e^- \to e^+ e^- \gamma$ has a very large cross section and can be an
important background if the electron (positron) is misidentified as a pion and the two
additional photons originate from split or accidental clusters.
$e^+ e^- \to \gamma \gamma$ has also a large cross section and may originate background
in case of photon conversions and there are split or accidental clusters.
Beside these, $\eta \gamma$ production with $\eta$ decaying to $\pi^+ \pi^- \gamma$ or to $3 \pi^0$ should be discriminated from the $\pi^+ \pi^- \pi^0$ signal by the number of prompt neutral clusters.

\subsection{Reconstruction of $\eta\gamma\to\pi^+ \pi^- \pi^0 \gamma$ events}

\begin{figure}[tbp]
 \centering
 \includegraphics[width = 8cm, height =8.5cm]{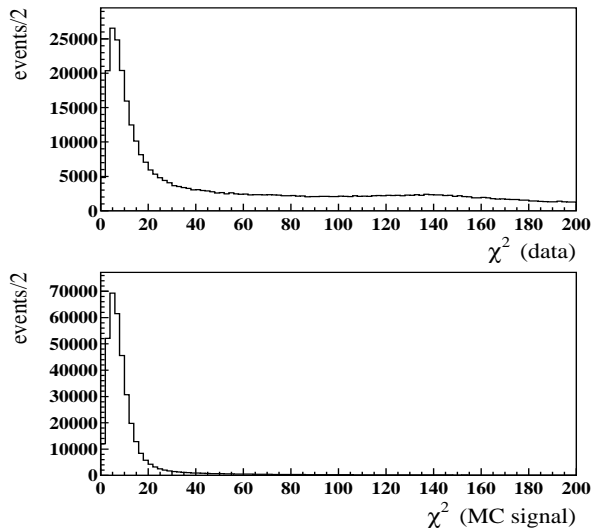}
 \caption{Distribution of $\chi^2$ for the data (top) and MC signal events (bottom).}
 \label{f:CHI2 KINFIT}
\end{figure}

The identification of the $\pi^0$ meson follows the procedure described in Section~\ref{s:RECO CHARGED}. 
No cut is applied to the value of $\chi^2_{\gamma \gamma}$.  A kinematic fit is applied to the selected combination of three neutral prompt clusters and two tracks, with the assignment of the charged pion mass. The fit uses 15 variables, the energy $E_i$, time $t_i$ and cluster coordinates $x_i, y_i, z_i$ of the three clusters, and has 7 constraints, promptness of three clusters $t_i - r_i/c = 0$, energy and momentum conservation: $\sum_{i}E_{\gamma i} + E_{\pi^+} + E_{\pi^-} = \sqrt{s}$ and $\sum_{i} \vec{p}_{\gamma i} + \vec{p}_{\pi^+} + \vec{p}_{\pi^-} = \vec{p}_{e^+ e^-}$. The track momenta are not varied in the minimization procedure.
Figure~\ref{f:CHI2 KINFIT} shows the distribution of the $\chi^2$ of the kinematic fit for MC signal events and for data. In the following analysis we select events with $\chi^2 < 50$.

\begin{figure}[tbp]
\centering
 \includegraphics[width=8cm,height=8cm]{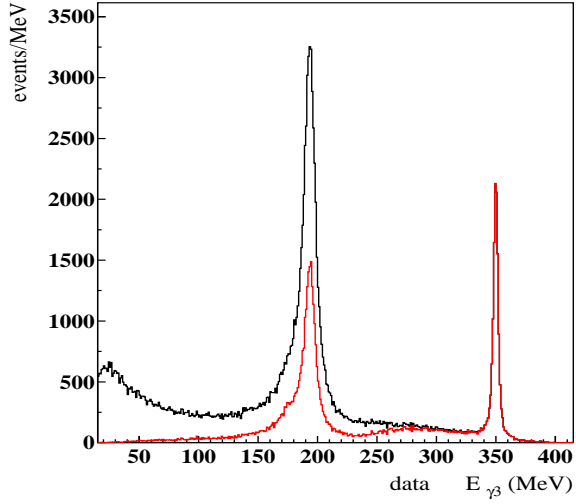}
 \caption{Distribution of the energy of the unpaired photon for data before (black) and after (red) the cut on the sum of the tracks momenta. The $e^+e^-\to\omega\gamma$ peak is clearly visible, with $E_\gamma=194$ MeV.}
\label{f:EGAMMA3}
\end{figure}

The background of $e^+ e^- \to e^+ e^- \gamma$ and $e^+ e^- \to \gamma \gamma$ is reduced using the $e$-$\pi$ likelihood estimator as described in Section~\ref{s:RECO CHARGED}, and requiring the angle between the two tracks to be $\alpha_{\pi \pi} < 160^{\circ}$, and the angle between any photon pair to be $\alpha_{\gamma \gamma} 
> 20^{\circ}$. The background of $e^+ e^- \to \eta \gamma \to 3 \pi^0 \gamma$ with photon conversion is reduced requiring the sum of the photon energies $\sum_{i}E_{\gamma i} < 660$ MeV. At this stage of the analysis, the residual background is dominated by the processes $e^+ e^- \to \pi^+ \pi^- \gamma$ and $e^+ e^- \to \pi^+ \pi^- \pi^0$ with split clusters, characterized by a neutral energy smaller than for the signal, and $e^+ e^- \to \omega \gamma \to  \pi^+ \pi^- \pi^0 \gamma$ characterized by the same final state as the signal. These backgrounds are reduced by requiring for the sum of the track momenta $|\vec{p}_+| + |\vec{p}_-| < 440$ MeV. The effect of these cuts is controlled by the distribution of the energy of the unpaired photon shown in Figure~\ref{f:EGAMMA3} where $E_{\gamma 3}$ is the value returned by the fit and has a resolution greatly improved by the good time and position resolution of the calorimeter. The peaks at the energies of the photon recoiling against the $\omega$ and the $\eta$ are clearly visible over a small background at $E_{\gamma 3} = 194$ MeV and $E_{\gamma 3} = 350$ MeV, respectively.

The selection efficiencies are evaluated with the MC simulation described in Section~\ref{s:MODEL} and are listed in Table~\ref{t:EFF_ETAG} for the signal and the most relevant background sources. 

\begin{table}[tbp]
 \centering
 \begin{tabular}{|c|c|c|}
\hline
 Final state & Selection & Global  \\
                  & efficiency (\%) & efficiency (\%)  \\
 \hline
 $\eta(\to \pi^+ \pi^- \pi^0)\gamma$ & 36.71 $\pm$ 0.02 & 28.68 $\pm$ 0.02 \\
 \hline
 $\pi^+ \pi^- \pi^0 \gamma$ & 6.08 & 1.19 \\
 $\omega \pi^0$ & 19.80 & 1.07 \\
 $\eta(\to \pi^+ \pi^- \gamma) \gamma$ & 0.723 & 0.069 \\
 $\eta (\to \mbox{neutral})\gamma$ & 0.111 & 0.002 \\
 \hline
  \end{tabular}
 \caption{Selection efficiency, in \%, for the signal and the most relevant backgrounds. The column Selection includes the efficiency of the trigger, the background filter  
and the data filter described in Section~\ref{s:DATASAMPLE}. }
 \label{t:EFF_ETAG}
\end{table}

\subsection{Evaluation of the cross section}

\begin{figure}[tbp]
 \includegraphics[width=8cm,height=8cm]{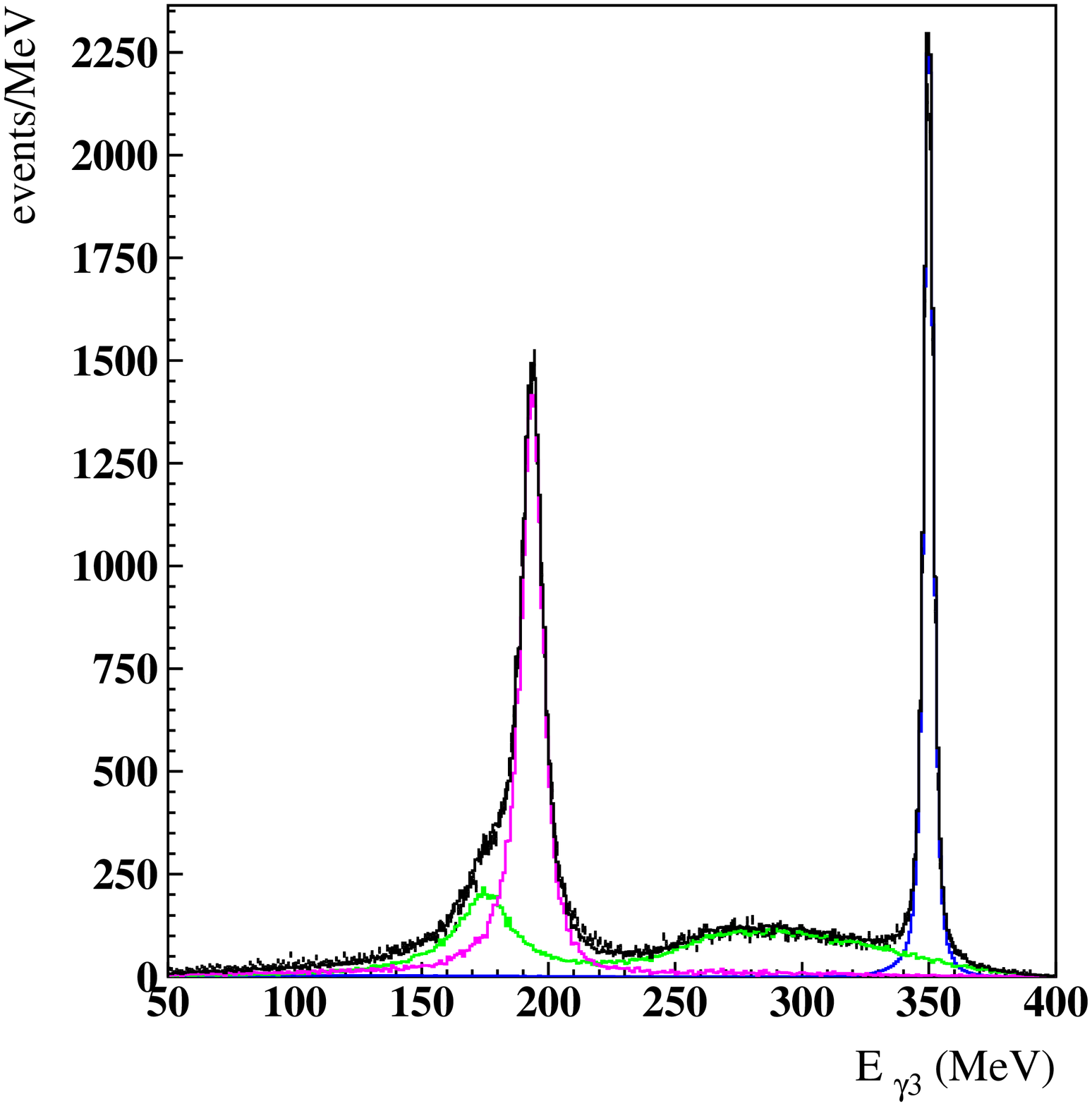}
\hspace{-0.5cm}
 \includegraphics[width=8cm,height=8cm]{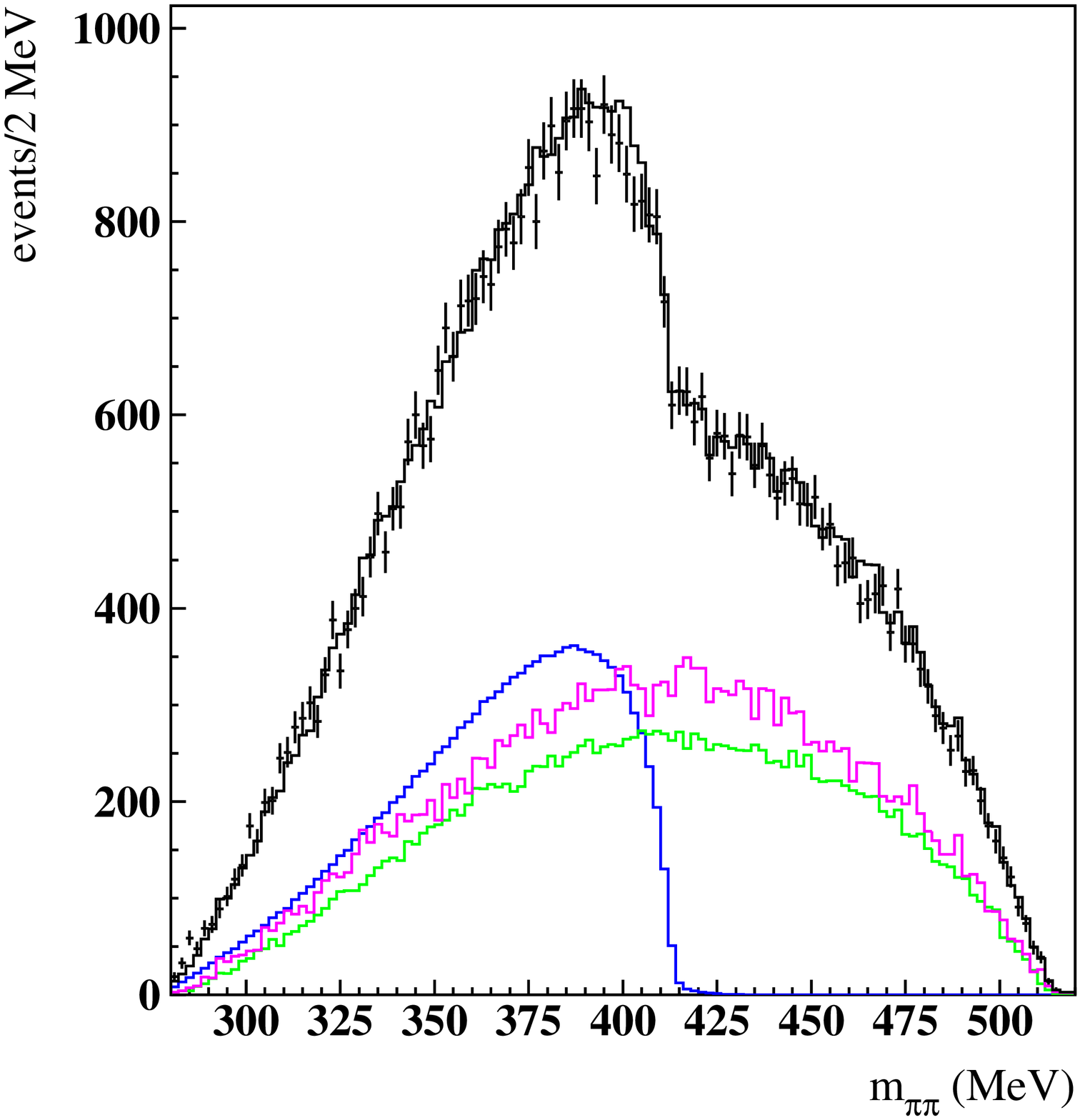}
 \caption{Projections of the 2-dimensional fit. Left: distribution of the energy of the unpaired photon. 
Right: distribution of the invariant mass $m_{\pi^+\pi^-}$. The contribution of the signal $e^+e^-\to\eta\gamma$ is blue, 
$e^+e^-\to\omega\pi^0$ is green and $e^+e^-\to\pi^+\pi^-\pi^0\gamma$ is purple.}
\label{f:EG3vsMPIPI}
\end{figure}

The number of signal events is derived with a 2-dimensional fit to the data. The distributions used to discriminate the signal from background are the energy of the unpaired photon and the invariant mass of the two charged pions in the interval 50 MeV $< E_{\gamma 3}<$ 400 MeV and 280 MeV $ < m_{\pi \pi} <$ 520 MeV that contains 55150 events. The fit to the data is done using the simulated shapes for the signal and backgrounds and the weights are left free. 
The projections of the $E_{\gamma 3}\times m_{\pi \pi}$ distribution are shown in Figure~\ref{f:EG3vsMPIPI} for the data and the backgrounds weighted by their relative factors returned by the fit. The result of the fit gives $13536 \pm 121$ signal events resulting in a cross section $\sigma(e^+ e^- \to \eta \gamma \to \pi^+\pi^-\pi^0\gamma) =  (194.7 \pm 1.8_{\mbox{stat}})$ pb. 

The only relevant backgrounds are from $e^+ e^- \to \pi^+ \pi^- \pi^0 \gamma$ and $e^+ e^- \to \omega \pi^0\to\pi^+\pi^-\pi^0\pi^0$. 
The distributions of the signal and $e^+ e^- \to \pi^+ \pi^- \pi^0 \gamma$ are well reproduced both in shape and relative normalization, while the fraction of $\omega \pi^0$ events results slightly higher than expected. If the measured value and its error, $\sigma(e^+ e^- \to \omega \pi^0\to\pi^+\pi^-\pi^0\pi^0) = (5.72 \pm 0.05)$ nb~\cite{K:omegapi0}, are introduced as a constraint, the fit returns a value 1.36\% higher for the $e^+ e^- \to \eta \gamma$ cross section. This difference is accounted for in the systematic error.

Other contributions to the systematic error are evaluated by varying the analysis cuts by the r.m.s. width of the distributions of each variable, $\chi^2$, $\alpha_{\pi \pi}$, $\alpha_{\gamma \gamma}$, $|\vec{p}_+| + |\vec{p}_-|$, 
accounting for their correlation. This results in a relative error of $\pm 1.45\%$ and  $\sigma(e^+ e^- \to \eta \gamma\to\pi^+\pi^-\pi^0\gamma) =  (194.7 \pm 1.8_{\mbox{stat}} \pm 2.8_{\mbox{syst}})$ pb. Using the branching fraction for $\eta \to \pi^+ \pi^- \pi^0$, we derive 
\begin{equation}
\sigma(e^+ e^- \to  \eta \gamma) = (856 \pm 8_{\mbox{stat}} \pm 12_{\mbox{syst}} \pm 11_{\mbox{\footnotesize{BR}}}) \mbox{\ pb} 
\label{eq:SIGMA ETAG}
\end{equation}
This value, obtained from a direct measurement, agrees well with the value~(\ref{eq:SIGMA ETAG000}) obtained from the analysis of $\gamma \gamma\to\eta\to 3\pi^0$. The result interpolates well with the measurements of the SND experiment~\cite{SND:etagamma} and has a better precision.

\section{Summary}
The cross section $\sigma(e^+ e^- \to e^+ e^- \eta)$ has been measured at $\sqrt{s} = 1$ GeV with the KLOE detector based on an integrated luminosity of 0.24 fb$^{-1}$. The $\eta$ mesons are selected using the two decays $\eta \to \pi^+ \pi^- \pi^0$ and $\eta \to \pi^0 \pi^0 \pi^0$ that exploit in a complementary way the tracking and the calorimeter measurements. Many background processes are considered, the most relevant being $e^+ e^- \to \eta \gamma$ when the photon is emitted at small polar angles and escapes detection. 
As a consistency check, we have measured the cross section for $e^+e^-\to\eta\gamma$ in two independent ways, the two values agree well with each other and we derive $\sigma(e^+ e^-\to\eta\gamma) = (856 \pm 8_{\mbox{stat}} \pm 16_{\mbox{syst}})$ pb. This value interpolates well previous measurements by the SND experiment and is more precise. The cross section for $e^+ e^- \to e^+ e^- \eta$ is obtained independently for the two $\eta$ decay modes with a 2-dimensional fit to the squared missing mass and the $\eta$ momentum projections. Combining the two measurements we obtain $\sigma(e^+ e^- \to e^+ e^- \eta) = (32.72 \pm 1.27_{\mbox{stat}} \pm  0.70_{\mbox{syst}})$ pb. This value is used to extract the partial width $\Gamma(\eta \to \gamma \gamma) = ( 520 \pm 20_{\mbox{stat}} \pm 13_{\mbox{syst}})$ eV. This is in agreement with the world average of $(510 \pm 26)$ eV and is the most precise measurement to date.


\acknowledgments
We wish to thank Fulvio Piccinini and Antonio Polosa for the countless support with the
Monte Carlo code for the signal generation and for enlightening discussions. 
We warmly thank our former KLOE colleagues for the access to the data collected during the KLOE data taking campaign.
We thank the DA$\Phi$NE team for their efforts in maintaining low background running conditions and their collaboration during all data taking. We want to thank our technical staff: 
G.F. Fortugno and F. Sborzacchi for their dedication in ensuring efficient operation of the KLOE computing facilities; 
M. Anelli for his continuous attention to the gas system and detector safety; 
A. Balla, M. Gatta, G. Corradi and G. Papalino for electronics maintenance; 
M. Santoni, G. Paoluzzi and R. Rosellini for general detector support; 
C. Piscitelli for his help during major maintenance periods. 
This work was supported in part by the EU Integrated Infrastructure Initiative Hadron Physics Project under contract number RII3-CT- 2004-506078; by the European Commission under the 7th Framework Programme through the `Research Infrastructures' action of the `Capacities' Programme, Call: FP7-INFRASTRUCTURES-2008-1, Grant Agreement No. 227431; by the Polish National Science Centre through the Grants No. 0469/B/H03/2009/37, 0309/B/H03/2011/40, DEC-2011/03/N/ST2/02641, 2011/01/D/ST2/00748 and by the Foundation for Polish Science through the MPD programme and the project HOMING PLUS BIS/2011-4/3.

\end{document}